\documentclass[10pt,twocolumn,aps,pra,superscriptaddress,showpacs,tightenlines,pdflatex,longbibliography]{revtex4-2}

\usepackage{tikz}
\usetikzlibrary{quantikz2}
\usepackage{amsthm}
\usepackage{amsmath}            %serve per le subequazioni
\usepackage{amssymb}            %serve per il simbolo "marchio registrato", \circledR
\usepackage{mathtools}
\usepackage{graphicx}           %serve per le figure eps, ps etcyy
\graphicspath{ {./images/} }
\usepackage[caption=false]{subfig}
\usepackage{multirow}

\usepackage{xcolor}
\usepackage{bm}
\usepackage{tcolorbox}
\usepackage{enumitem}
\usepackage[colorlinks,citecolor=blue,urlcolor=blue,linkcolor=blue,bookmarks=false,hypertexnames=true]{hyperref} 

\newcommand{\tr}{\text{Tr}}

\usepackage[normalem]{ulem}

\newcommand{\comt}[1]{}

\begin{document}
\title{Detecting basis-dependent hardware errors through spatio-temporal quantum steering}

\author{Hsiang-Wei Huang}
\affiliation{Department of Physics, National Cheng Kung University, 701401 Tainan, Taiwan}
\affiliation{Center for Quantum Frontiers of Research \& Technology (QFort), National Cheng Kung University, Tainan 701401, Taiwan}

\author{Kuo-Feng Chiu}
\affiliation{Department of Physics, National Cheng Kung University, 701401 Tainan, Taiwan}
\affiliation{Center for Quantum Frontiers of Research \& Technology (QFort), National Cheng Kung University, Tainan 701401, Taiwan}

\author{Yi-Te Huang}
\affiliation{Department of Physics, National Cheng Kung University, 701401 Tainan, Taiwan}
\affiliation{Center for Quantum Frontiers of Research \& Technology (QFort), National Cheng Kung University, Tainan 701401, Taiwan}

\author{Jhen-Dong Lin}
\affiliation{Department of Physics, National Cheng Kung University, 701401 Tainan, Taiwan}
\affiliation{Center for Quantum Frontiers of Research \& Technology (QFort), National Cheng Kung University, Tainan 701401, Taiwan}

\author{Tse-Ming Chen}
\affiliation{Department of Physics, National Cheng Kung University, 701401 Tainan, Taiwan}
\affiliation{Center for Quantum Frontiers of Research \& Technology (QFort), National Cheng Kung University, Tainan 701401, Taiwan}

\author{Yueh-Nan Chen}
\email{yuehnan@mail.ncku.edu.tw}
\affiliation{Department of Physics, National Cheng Kung University, 701401 Tainan, Taiwan}
\affiliation{Center for Quantum Frontiers of Research \& Technology (QFort), National Cheng Kung University, Tainan 701401, Taiwan}
\affiliation{Physics Division, National Center for Theoretical Sciences, Taipei 106319, Taiwan}

\begin{abstract}

Spatio-temporal quantum steering provides a framework for benchmarking the nonclassicality of general quantum state transfer processes. A central diagnostic is the no-signaling-in-time (NSIT) condition, whose violation can indicate basis-dependent hardware errors. However, finite measurement statistics may also yield apparent violations, thereby obscuring the detection of basis-dependent hardware errors. To address this, we construct a statistical hypothesis test under the null hypothesis that NSIT violations arise solely from statistical fluctuations. Combining the statistical properties of NSIT violation under the null hypothesis with Chebyshev’s inequality, we obtain a distribution-free upper bound on the $p$-value without parametric assumptions.
We apply this method to two examples. For a single-qubit state-transfer experiment on a superconducting processor, we observe several instances that the NSIT violation is observed and the null hypothesis is simultaneously rejected by a small $p$-value, providing statistical evidence of basis-dependent hardware errors. For a seven-qubit Hayden–Preskill teleportation protocol on IonQ devices, the null hypothesis is also rejected even when the average fidelity exceeds the classical threshold, while the associated nonclassicality measure vanishes. Our results highlight the necessity of statistical hypothesis testing for detecting basis-dependent errors in near-term quantum devices.

\end{abstract}

\maketitle

\section{Introduction}\label{sec:intro}
The concept of quantum steering was first introduced by Schr\"{o}dinger~\cite{Schrodinger_1935}, who showed that Alice can remotely steer Bob's quantum state through her choice of measurement basis on one part of an entangled pair. With proper mathematical formulation, quantum steering~\cite{wiseman2007, Cavalcanti2017, Uola2020, Xiang2022} has been recognized as a valuable resource for various semi-device-independent quantum information tasks, including subchannel discrimination~\cite{Piani2015, Sun2018}, quantum key distribution~\cite{Bartkiewicz2016}, and quantum metrology~\cite{Yadin2021} (see also Refs.~\cite{Uola2020} for comprehensive reviews).

Recently, a spatio-temporal generalization of quantum steering, known as spatio-temporal steering (STS), has been proposed~\cite{Chen_2017, Chen2014, Li2015, Ku_2018}. Unlike conventional fidelity-based metrics, STS can reveal the nonclassical features of general quantum state-transfer (QST) processes~\cite{Yite_2021}. In this framework, violations of the no-signaling-in-time (NSIT) condition~\cite{Simon_2001, Gisin2020, Rybotycki2025} quantify temporal signaling between state preparation and subsequent measurements. In practical quantum devices, such signaling may originate from basis-dependent hardware imperfections, including measurement cross-talk~\cite{Silman_2013, Sarovar_2020, S.Seo_2022, Xie_2010, Rudinger_2021, Tabia_2025}, context-dependent errors~\cite{Rudinger2019, Debroy2023, Ai2025, Seif2024, Marton2023, Greenbaum_2018}, and gate-dependent errors~\cite{Epstein2014, Wallman2018, Merkel2021, Brieger2025, Chen2021, Magesan2012}.

In realistic quantum experiments, however, measurements are performed with a finite number of shots, and the statistical fluctuations alone may also lead to apparent violations of no-signaling conditions~\cite{Engineer_2025}. It is therefore essential to determine whether an observed violation arises purely from finite sampling or instead signals the presence of systematic hardware imperfections. Similar questions have been widely studied in the literature on quantum characterization, verification, and validation~\cite{blumekohout2025, BlumeKohout2020, Proctor2019, Proctor2022, Magesan2011, Mezher2022, Knill2008, Alexander2016, Merkel2013, greenbaum2015, Nielsen2021}, where statistical hypothesis testing~\cite{Evans2004} is used to provide the statistical evidence against the null hypothesis~\cite{Elkouss2016, Shalm2015}.

In this work, we aim to provide a hypothesis-testing framework for distinguishing statistical fluctuations from intrinsic basis-dependent errors under the independent and identically distributed (i.i.d.) assumption. The null hypothesis is that any NSIT violation arises purely from statistical fluctuation. Specifically, by combining the statistical properties of NSIT violation under the null hypothesis with Chebyshev's inequality~\cite{Vershynin_2026}, we derive an upper bound on the $p$-value for a given number of measurement shots. The constructed upper bound requires no assumption about the underlying distribution and is well-suited to the finite-shot regime of current quantum devices. This enables the null hypothesis to be rejected when the $p$-value is bounded below the chosen significance level $\alpha$. Additionally, we characterize the non-classicality of the QST process by estimating the spatio-temporal steering robustness (STSR) of NSIT-compatible assemblage reconstructed from the empirical assemblage through the non-signaling model algorithm~\cite{Engineer_2025}.

We apply this method to two examples. The first one is a single-qubit QST task implemented on a superconducting processor, where the qubit is left idle for a duration $t$ after state preparation. We observe a consistent violation of the NSIT condition. By applying the hypothesis testing to the observed NSIT violation, we observe several instances in which the null hypothesis is rejected by a sufficiently small $p$-value. This provides the statistical evidence for a basis-dependent error. In the second example, we consider a seven-qubit Hayden–Preskill teleportation protocol~\cite{Yoshida_2019} designed to certify the underlying quantum information scrambling based on the average teleportation fidelity. We implement this protocol on cloud-trapped-ion quantum computers provided by IonQ (IonQ Aria1 and IonQ Harmony). In particular, for the IonQ Harmony device, although the average teleportation fidelity surpasses the classical threshold, the null hypothesis is rejected by a $p$-value strictly less than $\alpha$. Additionally, the associated STSR vanishes after applying the non-signaling model algorithm. This suggests that, despite the teleportation fidelity exceeding the classical bound, it remains questionable whether the experiment verifies the device's ability to generate genuine quantum information scrambling.

The remainder of the paper is structured as follows. In Sec.~\ref {sec: theory}, we briefly review the theoretical framework of quantum steering and introduce how hypothesis testing can be used for identifying the basis-dependent hardware error. In Sec.~\ref{sec: experiment 1}, we employ our method to a single-qubit QST task implemented on a superconducting quantum processor, in which we identify the statistical evidence of basis-dependent hardware error in the selected qubit. In Sec.~\ref{sec: exp2}, we further employ our method on the Hayden–Preskill teleportation protocol conducted on the cloud IonQ quantum computer. The results from IonQ Harmony suggest that using teleportation fidelity as a non-classical metric may be questionable. Finally, we summarize our results in Sec.~\ref {sec: conclusion}.

\section{Theoretical Framework}\label{sec: theory}
In this section, we present the theoretical framework for using hypothesis testing to characterize the potential basis-dependent hardware errors in a given QST experiment.
\subsection{Non-classicality of quantum state transfer through spatio-temporal steering}
Before discussing the basis-dependent error, we revisit the QST experiment in the spatio-temporal steering scenario~\cite{Yite_2021}. 

\begin{figure*}[!hbpt]
    \includegraphics[width=\textwidth]{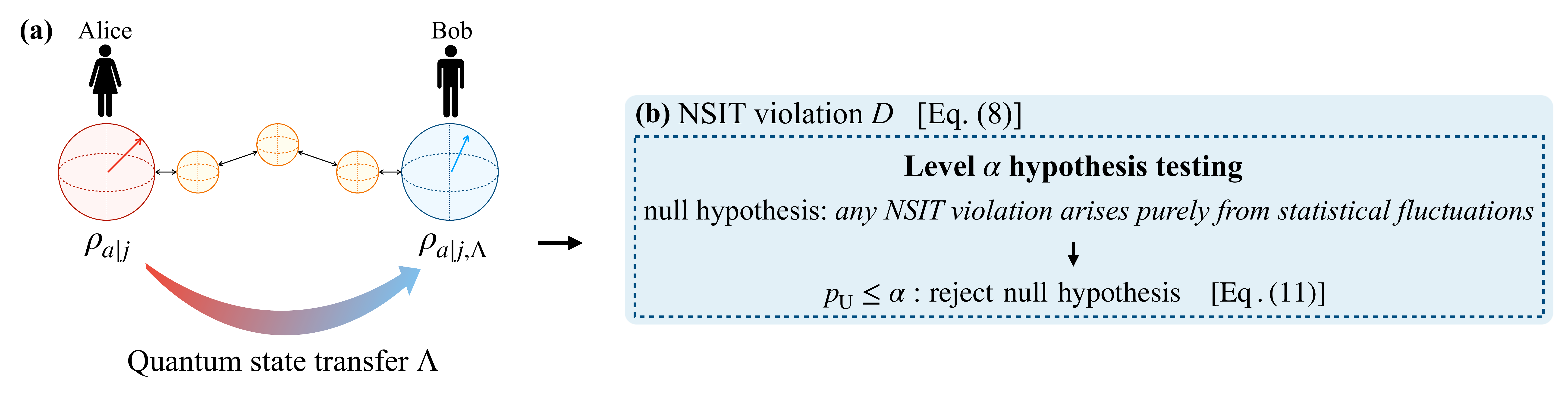}
    \caption{The illustration of our result. (a) In the STS framework introduced in Sec.~\ref {sec: theory}, Alice prepares the state $\rho_{a|j}$ and transmits it to Bob via the QST process $\Lambda$ with different indices $a$ and $j$. Then Bob collects the received state $\rho_{a|j,\Lambda}$ to construct the assemblage $\{\tilde{\rho}_{a|j,\Lambda}\}$, which characterizes the non-classicality of the QST process. (b) We investigate the presence of basis-dependent hardware errors in the QST process by performing a level $\alpha$ hypothesis testing on the NSIT violation $D$ of $\{\tilde{\rho}_{a|j,\Lambda}\}$. If the resulting \textit{p}-value does not exceed $\alpha$, the null hypothesis that any NSIT violation arises purely from statistical fluctuation is rejected, suggesting that it may instead stem from a hardware-dependent mechanism. 
    % (c) We can employ the NSMA in Eq.~\eqref{eq: NSMA} to map $\{\tilde{\rho}_{a|j,\Lambda}\}$ to the closest NSIT-compatible assemblage, and certify the genuine non-classicality of $\Lambda$ via the corresponding STSR. The QST process $\Lambda$ is deemed non-classical whenever STSR $>0$. 
    }
    \label{fig: illustration}
\end{figure*}

A general QST task aims to transmit an arbitrary quantum state $\rho_{\mathrm A}$ from a sender (Alice) to a receiver (Bob) through a quantum channel $\Lambda$, such that
\begin{equation}\label{eq:QST}
    \rho_{\mathrm B} = \Lambda(\rho_{\mathrm A}),
\end{equation}
where $\rho_{\mathrm B}$ denotes the output state received by Bob. To characterize the non-classicality of the process, we adopt the STS approach proposed in Ref.~\cite{Yite_2021}, as illustrated in Fig.~\ref{fig: illustration} (a).

Throughout this work, we focus on the STS for qubit systems, where Alice prepares quantum states chosen from the eigenbases of the Pauli operators $\{\sigma_x, \sigma_y, \sigma_z\}$. Specifically, for each basis $j \in \{x,y,z\}$, she prepares the eigenstates
\begin{equation}\label{eq: state def}
    \rho_{a|j} = \frac{1}{2}\left( \openone + (-1)^a~ \sigma_j \right) \text{~~where~~} a \in \{0,1\},
\end{equation}
and sends them through the QST process $\Lambda$. The output states received by Bob are then given by
\begin{equation}
    \rho_{a|j,\Lambda} = \Lambda(\rho_{a|j}).
\end{equation}
We further define the spatio-temporal steering assemblage, which is characterized by a collection of subnormalized states:
\begin{equation}
    \tilde{\rho}_{a|j,\Lambda} = p(a|j)\,\rho_{a|j,\Lambda},
\end{equation}
where $p(a|j)$ denotes the preparation probability. For simplicity, we assume uniform preparation, $p(a|j)=1/2$, under which the no-signaling-in-time (NSIT) condition is satisfied:
\begin{equation}
    \sum_a \tilde{\rho}_{a|j,\Lambda}
    =
    \sum_a \tilde{\rho}_{a|j',\Lambda}
    =
    \frac{1}{2}\openone,
    \qquad \forall\, j \neq j'.
\end{equation}

According to Ref.~\cite{Yite_2021}, a QST process is considered classical if it can be described by a measure-and-prepare channel~\cite{Horodecki_2003}. In this case, the resulting assemblage is unsteerable and admits a classical hidden-state (HS) model:
\begin{equation}\label{eq:LHS}
    \tilde{\rho}^{\mathrm{HS}}_{a|j}
    =
    \sum_{\lambda}
    p(\lambda)\, p(a|j,\lambda)\, \rho_{\lambda},
    \qquad \forall a,j,
\end{equation}
where $\lambda$ is a classical hidden variable distributed according to $p(\lambda)$, $p(a|j,\lambda)$ is a classical response function, and $\rho_\lambda$ are predetermined quantum states. Consequently, the non-classicality of the QST process can be certified whenever the generated assemblage is steerable, i.e., when no such hidden-state decomposition exists.

To quantify this steerability, we employ the spatio-temporal steering robustness (STSR), defined as
\begin{equation}
\begin{aligned}
    &\mathrm{STSR}(\{\tilde{\rho}_{a|j,\Lambda}\}_{a,j})
    = \min_{r,\{\tilde{\tau}_{a|j}\},\{\tilde{\rho}^{\mathrm{HS}}_{a|j}\}} r\\
    \text{s.t. }&\frac{1}{1+r}\tilde{\rho}_{a|j,\Lambda}
    +
    \frac{r}{1+r}\tilde{\tau}_{a|j}
    =
    \tilde{\rho}^{\mathrm{HS}}_{a|j},
    ~\forall a,j.
\end{aligned}
\end{equation}
This optimization problem can be efficiently solved using a semidefinite program~\cite{Chen_2017}. Operationally, the STSR quantifies the minimal amount of noise $\{\tilde{\tau}_{a|j}\}$ that must be mixed with the given assemblage to render it unsteerable. Therefore, a QST process is certified as non-classical when both $\mathrm{STSR} > 0$ and the NSIT condition is satisfied~\cite {Yite_2021}.

\subsection{Hypothesis testing for identifying the intrinsic basis-dependent hardware errors}
As shown in Ref.~\cite{Yite_2021}, the NSIT condition can be violated in real-world implementations, where the degree of violations can be quantified using a trace-distance-based measure~\cite{Yite_2021, Ku_2018}, which is defined as
\begin{equation}
    D(\{\tilde{\rho}_{a|j,\Lambda}\}_{a,j}) = \max_{j,j'}\frac{1}{2} 
    \left|\left|  
        \sum_a\tilde{\rho}_{a|j, \Lambda} - \sum_{a'}\tilde{\rho}_{a'|j', \Lambda}
    \right|\right|_1.
\end{equation}
Here, $D=0$ if and only if the NSIT condition is satisfied. 

Within the operational framework of QST, such violations may originate from at least two distinct types of basis-dependent hardware imperfections. The first type arises from state-preparation errors, where the implemented input states deviate from their intended target states in a basis-dependent manner. This effect can be modeled by introducing a basis-dependent noisy channel $\mathcal{E}_{a|j}$ such that the actual prepared state $\varrho_{a|j}$ is given by
\begin{equation}
\varrho_{a|j} = \mathcal{E}_{a|j}(\rho_{a|j}).
\end{equation}
This type of imperfection is commonly referred to as a gate-dependent error~\cite{Epstein2014, Wallman2018}.

The second type originates from the context dependence of the QST process itself~\cite{Rudinger2019}. In this scenario, the effective channel governing the QST depends on the preparation basis, such that the nominal channel $\Lambda$ is replaced by a basis-conditioned channel $\Lambda_{a|j}^*$. This reflects a setting-dependent modification of the implemented process rather than an imperfection in state preparation alone.

In addition, statistical fluctuations can also induce apparent NSIT violations, since the transferred states $\rho_{a|j,\Lambda}$ are reconstructed via quantum state tomography with a finite number of measurement shots. To determine whether an observed violation $D$ originates from hardware imperfections rather than statistical fluctuations alone, we formulate a level-$\alpha$ hypothesis test under the assumption of i.i.d. sampling. The corresponding null hypothesis $\mathcal{H}_0$ is stated as follows: \textit{any observed NSIT violation arises solely from statistical fluctuations.}
To characterize the statistical behavior under $\mathcal{H}_0$, we perform numerical simulations using the Amazon Braket simulator~\cite{braket}. Since the simulator is an ideal error-free channel, the resulting distribution of $D$ reflects only finite-sampling statistical fluctuations.

Further, we consider the test statistic $T$, a function of data for quantifying the deviation from the null hypothesis $\mathcal{H}_0$, defined by
\begin{equation}\label{eq: statistic}
    T(D) = \frac{|D - \mu|}{\sigma},
\end{equation}
where the mean value $\mu$ and standard deviation $\sigma$ are estimated from 100,000 samples. The plausibility of the null hypothesis can then be estimated from the \textit{p}-value of $T$, which represents the tail probability for the observed value $t$ of $T$ if the null hypothesis $\mathcal{H}_0$ holds~\cite{lehmann2005}, that is
\begin{equation}
    p\text{-}\mathrm{value} = \mathrm{Prob}(T\geq t~| \mathcal{H}_0~\mathrm{holds}).
\end{equation}
If the \textit{p}-value falls below the significance level $\alpha$, for which we adopt the conventional choice $\alpha=0.05$, we consider it statistically significant to reject the null hypothesis. 

Instead of computing an exact \textit{p}-value under a fully specified distribution, we apply Chebyshev's inequality~\cite{Vershynin_2026} to derive a distribution-independent \textit{p}-value upper bound $p_\mathrm{U}$:
\begin{equation}\label{eq: bound}
   p\text{-}\mathrm{value} \leq \frac{1}{t^2} \eqqcolon p_\mathrm{U}.
\end{equation}
From the considered significance level $\alpha=0.05$, the above equation implies that any observed deviation from $\mu$ exceeding $\sqrt{20} \sigma$ provides statistical evidence against the null hypothesis. Although adopting Chebyshev's inequality sacrifices the precision in the \textit{p}-value, it guarantees a valid level-$\alpha$ test without invoking Gaussianity or other parametric assumptions, making it robust across distributions~\cite{Vershynin_2026}.

To characterize the genuine non-classicality of the QST process when the empirical assemblage $\{\tilde{\rho}_{a|j,\Lambda}\}_{a,j}$ violates the NSIT condition, we employ the one-sided device independent non-signaling model algorithm proposed in Ref.~\cite{Engineer_2025} to find the closest NSIT-compatible assemblage. This can be achieved by solving the following semidefinite program :
\begin{equation}\label{eq: NSMA}
    \begin{aligned}
     r &= \min_{\{\tilde{\tau}_{a|j}\}_{a,j}} \tr(\sum_{a}\tilde{\rho}_{a|j,\Lambda} + \tilde{\tau}_{a|j})-1 ,\\
    \text{s.t.}\quad \sum_{a}&\left[\tilde{\rho}_{a|j,\Lambda} + \tilde{\tau}_{a|j}\right] = \sum_{a'}\left[\tilde{\rho}_{a'|j',\Lambda} + \tilde{\tau}_{a'|j'}\right]  \quad\forall j,j'.
\end{aligned}
\end{equation}

Once the optimal solution is obtained, the closest NSIT-compatible assemblage $\{\tilde{\rho}_{a|j,\Lambda}^{\mathrm{NSIT}}\}$ can be reconstructed by 
\begin{equation}
    \tilde{\rho}_{a|j,\Lambda}^{\mathrm{NSIT}} = \frac{\tilde{\rho}_{a|j,\Lambda} + \tilde{\tau}^*_{a|j}}{1+r},
\end{equation}
where $\tilde{\tau}_{a|j}^*$ yields the optimal $r$ of Eq.~\eqref{eq: NSMA}. The genuine non-classicality is subsequently quantified by estimating its corresponding STSR. Throughout the experiments presented below, the non-signaling model algorithm is always applied prior to estimating the STSR in order to reflect the genuine non-classicality of the QST process.

\section{Example 1: Single-qubit QST on superconducting processor}\label{sec: experiment 1}

\begin{figure}
    \centering
    \includegraphics[width=0.8\columnwidth]{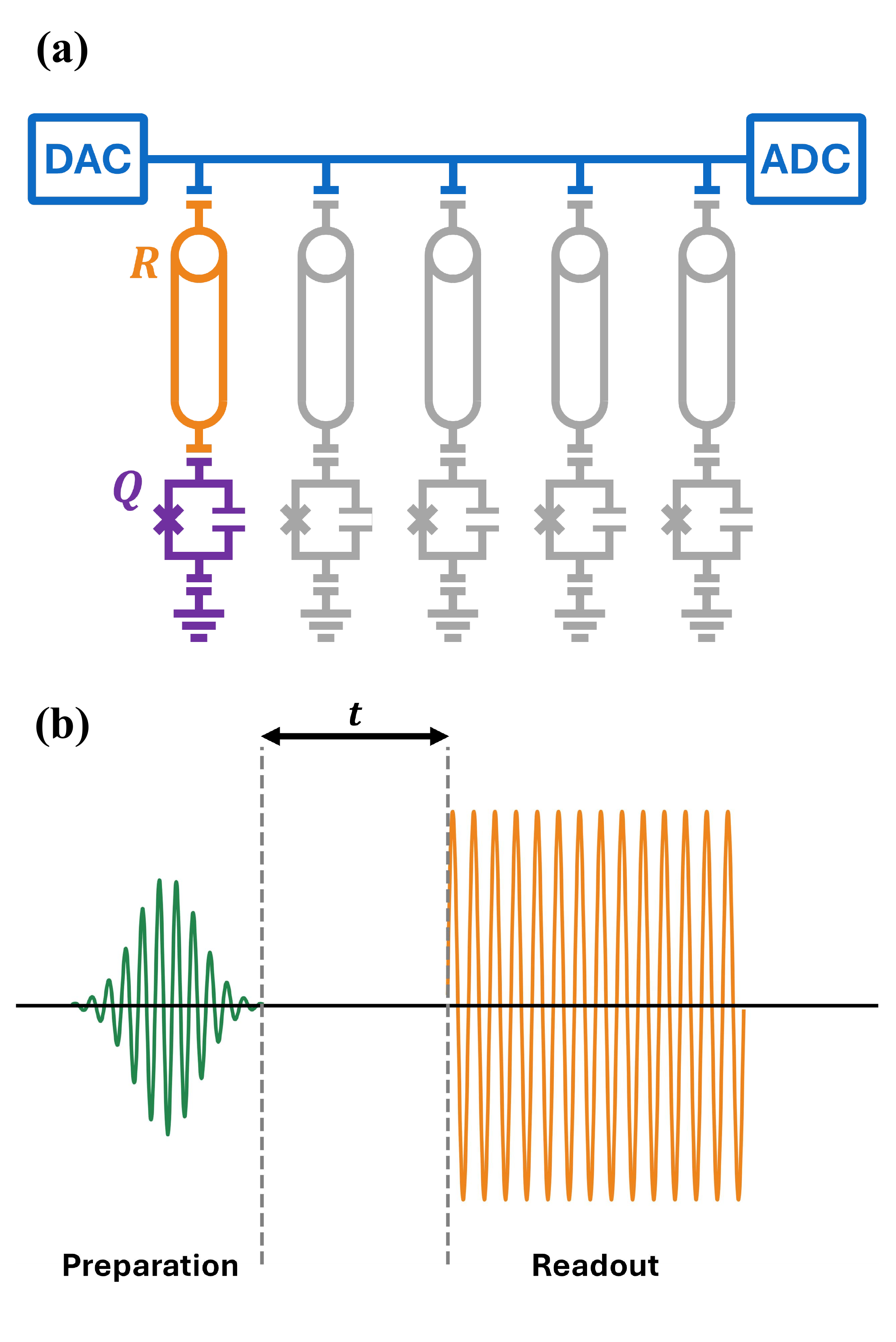}
    \caption{Device schematic and experimental pulse sequence. (a) Circuit diagram showing the experimental system of the five-qubit superconducting quantum processor. The active qubit ($Q$) under study and its readout resonator ($R$) are highlighted in color, whereas the other four qubits on the same processor are inactive during the single-qubit QST experiment. (b) Sequence of control and measurement pulses for the single-qubit QST experiment. The qubit is prepared to each state $\rho_{0|z}$, $\rho_{1|z}$, $\rho_{0|x}$, $\rho_{1|x}$, held for an idling time $t$, and subsequently measured.}
    \label{fig:QST_pulse}
\end{figure}
\begin{center}
\begin{figure}[!hbpt]
    \includegraphics[width=\columnwidth]{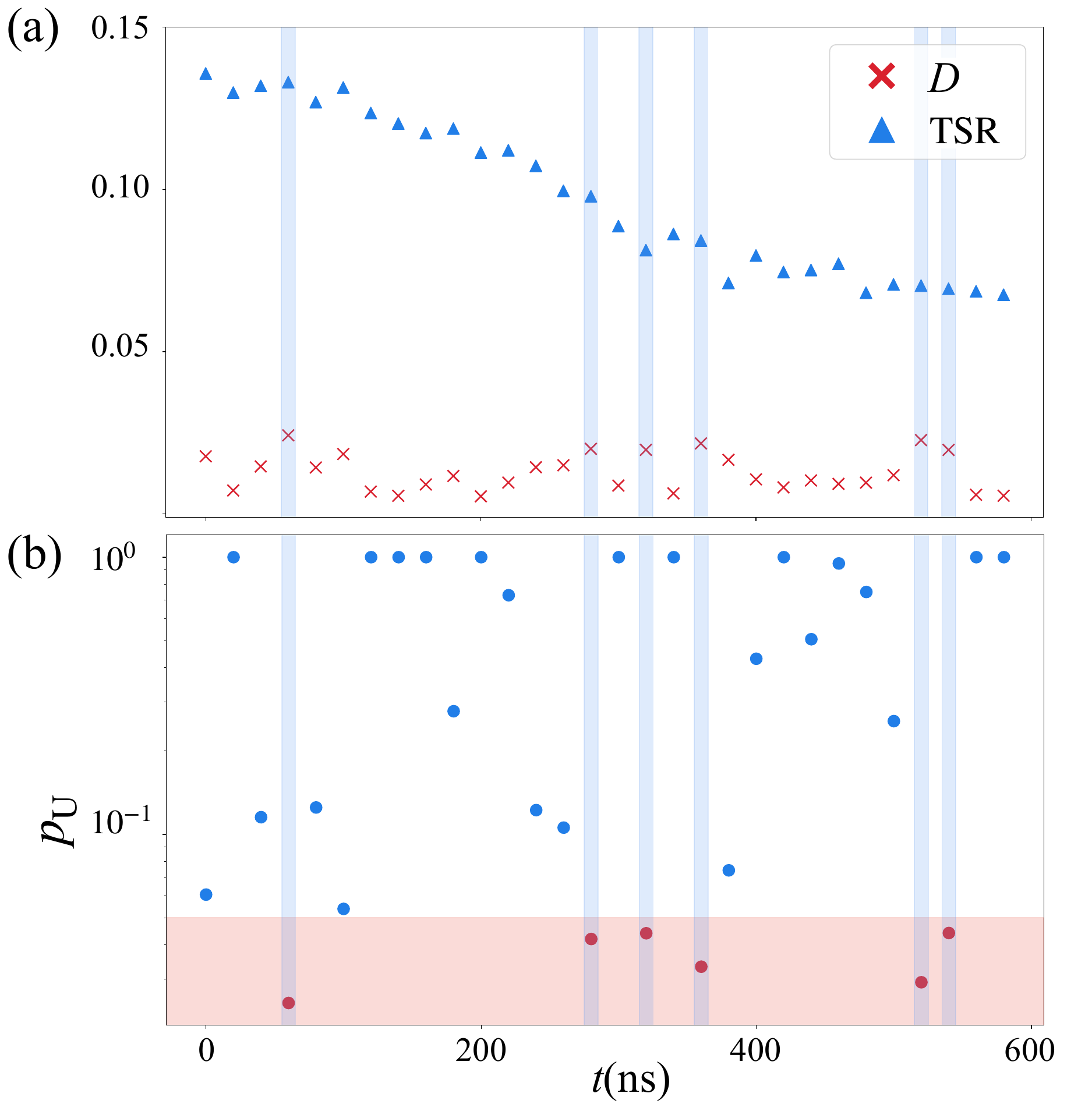}
    \caption{Single-qubit QST task on the superconducting quantum processor. (a) The non-classicality of QST captured by TSR and the degree of NSIT violation $D$ at different time points. (b) By evaluating $p_\mathrm{U}$ at each time point by Eq.~\eqref{eq: bound}, we find instances within the red shaded area provide the statistical evidence ($p_\mathrm{U} \leq 0.05$) of basis-dependent hardware errors.
    }
    \label{fig: exp1}
\end{figure}
\end{center}

In this section, we present a single-qubit QST experiment performed on a superconducting quantum processor. The initial steering assemblage $\{\rho_{a|j}\}_{a,j}$ defined in Eq.~\eqref{eq: state def} is prepared at time $t_0$. The qubit is then left idle, allowing the assemblage to evolve naturally to a later time $t$ under the intrinsic device dynamics. The transferred assemblage at time $t$, denoted as $\{\tilde{\rho}_{a|j, \Lambda}(t)\}_{a,j}$, is reconstructed via quantum state tomography. Since the protocol involves only a single qubit, the spatio-temporal steering scenario reduces to the temporal steering framework~\cite{Chen2014}. Consequently, the measured STSR can equivalently be interpreted as the temporal steering robustness (TSR)~\cite{Chen2014, Chen_2017}.

The experiment is carried out on a five-qubit superconducting quantum processor (see Fig.~\ref{fig:QST_pulse} and Appendix~\ref{appx: qubit layout} for details). Specifically, we select the qubit $Q$ to implement the single-qubit QST protocol. The idling time is varied from 0 to 600 ns. For each duration, quantum state tomography is performed with 10,000 measurement shots. As shown in Fig.~\ref{fig: exp1}(a), the resulting TSR decays over time due to decoherence, indicating the gradual loss of non-classicality in the QST process. 

The signature of NSIT violation at each time point is identified by a non-vanishing value of $D$. To assess whether the observed $D$ can be attributed purely to statistical fluctuations, we apply the proposed hypothesis test on the observed value of $D$. Simulations on the Amazon Braket simulator show that, under the null hypothesis, the mean value $\mu$ and standard deviation $\sigma$ of $D$ in this experiment are 0.0064 and 0.0028, respectively.

Using these statistical properties of the null hypothesis, we evaluate the test statistic $T$ and bound the \textit{p}-value at each time point by Eq.~\eqref{eq: bound}. We find that $p_\mathrm{U}\geq \alpha$ at most of the time points, indicating that statistical fluctuations cannot be excluded in those cases. Nevertheless, within the red shaded area of Fig.~\ref{fig: exp1}(b), a number of time points exhibit $p_\mathrm{U}\leq \alpha$, indicating that the null hypothesis is rejected. This suggests that statistical fluctuations alone are unlikely to fully account for the observed violation in those instances, and therefore provides statistical evidence for the presence of an intrinsic, basis-dependent hardware error in the selected qubit.

\section{Example 2: Hayden-Preskill teleportation protocol on cloud trapped ion quantum computers}\label{sec: exp2}
\begin{center}
    \begin{figure}[!hbpt]\label{protocol}
        \includegraphics[width=1.0\columnwidth]{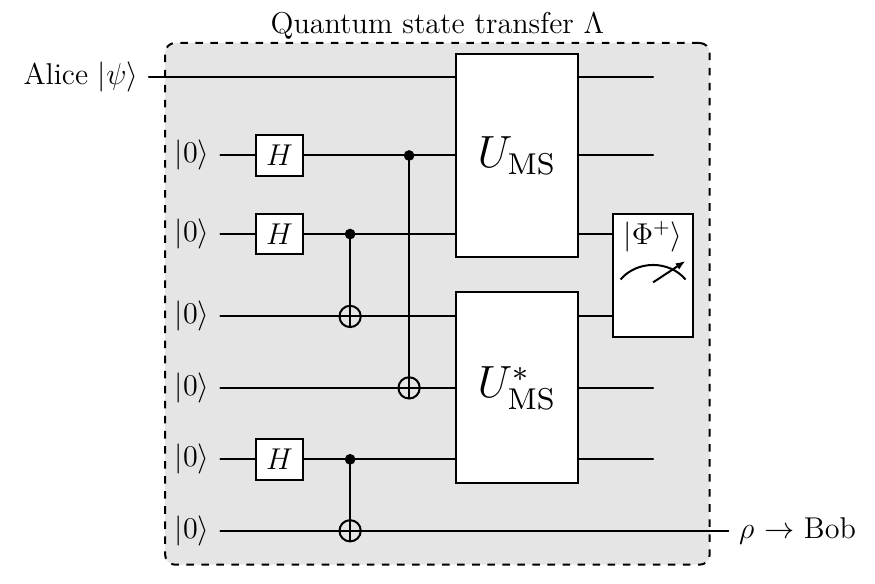}
        \caption{Teleportation circuit based on the Hayden-Preskill decoding protocol~\cite{Hayden_2007}. 
       The maximally scrambling unitary [$U_\text{MS}$ and its conjugate $U^*_\text{MS}$] encodes Alice's state (information) into the global system. After performing the Bell state measurement, Bob can decode Alice's state by post-selecting the outcome corresponding to the $|\Phi^+\rangle = (|00\rangle + |11\rangle) / \sqrt{2}$ state.
        %Two scramblers $U(t)$ and $U(t)^*$ scramble the information of Alice's state to the global system. After Bell measurement, Bob can decode Alice's state when the outcome of Bell measurement is in $|EPR\rangle = \frac{1}{2}(|00\rangle + |11\rangle)$ state.
        }
        \label{fig: protocol}
    \end{figure}
\end{center}
\begin{center}

\begin{table}
    \caption{
    Here, we show the results from different devices with the corresponding STSR, $D$, $p_{\mathrm{U}}$, and average teleportation fidelity $\mathcal{F}_{\textrm{avg}}$. Notably, the Harmony device's results show a signature of a basis-dependent hardware error.
    }    
    \label{table: scrambling rs}
        \vspace{5.5pt}
        \begin{tabular}{lccccc}
            \hline
            \hline
            Device ~~~~~~& $~~\mathrm{STSR}~~$ & ~~$D$~~ & ~~$\mathcal{F}_{\textrm{avg}}$~~ & $~~p_{\mathrm{U}}~~$\\
            \hline
            {Ideal Simulator}~~~~~~ & {0.269} & {0.030} & {0.999} &{$ 1$}\\
            {IonQ Aria1} & {0.128}& {0.049} & {0.906} & {$ 1$}\\
            
            {IonQ Harmony} & {0.000} & {0.095} &{0.742} & {$  0.04$}\\
            \hline
            \hline
        \end{tabular}
  \end{table}
\end{center}
In this section, we revisit the Hayden–Preskill teleportation experiment proposed in Refs.~\cite {Yoshida_2019, Landsman_2019, Hayden_2007}. The circuit implementation is depicted in Fig.~\ref{fig: protocol}, where Alice aims to teleport her quantum state $\ket{\psi}$ to Bob conditioned on the successful post-selection of the Bell state measurement. The three-qubit unitary gate $U_{\mathrm{MS}}$, whose circuit decomposition is shown in Appendix~\ref{appx: scrambler circuit}, is referred to as the maximally scrambling unitary. This gate generates a strong quantum information scrambling effect for a three-qubit system and enables perfect Alice-to-Bob state teleportation~\cite{Landsman_2019}.  

As reported in Ref.~\cite{Yoshida_2019}, the teleportation fidelity may be degraded due to the hardware imperfections such as the decoherence effect or the mismatch between $U_\text{MS}$ and $U^*_\text{MS}$~\cite{Yoshida_2019}. Nevertheless, it is believed that a genuine ‘‘quantum" scrambling effect can still be verified when the observed average teleportation fidelity surpasses the conventional classical threshold $\mathcal{F}_\text{cl}=2/3$.

However, based on our steering-based test, we find that teleportation fidelity alone may be insufficient to determine whether the underlying scrambling effect is genuinely quantum. Specifically, as we will demonstrate below, the average teleportation fidelity can exceed the classical threshold even when basis-dependent intrinsic hardware errors cannot be completely ruled out, thereby making the fidelity-based indicator unreliable.   

We experimentally implement the teleportation protocol on trapped-ion quantum computers provided by IonQ (IonQ Aria1 and IonQ Harmony~\cite{IonQ_Harmony}). Similarly, we prepare Alice's initial-state assemblage as described in Eq.~\eqref{eq: state def}. Then the steering assemblage received by Bob was tomographically reconstructed using 2000 shots, and the corresponding distribution of $D$ under the null hypothesis has the mean value $\mu=0.038$ and the standard deviation $\sigma=0.012$. The results for the experiments are summarized in Table~\ref{table: scrambling rs}.

The results of the IonQ Aria-1 device demonstrate high average teleportation fidelity $\mathcal{F}_\text{avg}\approx 0.9057$, with a low NSIT violation $D \approx 0.0491$. In addition, the \textit{p}-value upper bound $p_{\mathrm{U}}$ is unity in this case, meaning that the null hypothesis cannot be rejected and one cannot rule out that the NSIT violation may solely come from statistical fluctuations. Moreover, the STSR is robust and nonzero after mapping the empirical assemblage onto the NSIT-compatible assemblage, reflecting the genuine nonclassicality of the QST process. Therefore, it is plausible that this example demonstrates genuine quantum information scrambling.

In contrast, for the IonQ Harmony device, the teleportation fidelity is significantly lower, $\mathcal{F}_\text{avg}\approx 0.7418$, but still exceeds the classical threshold. However, the observed value of $D$ is unexpectedly large compared to the mean $\mu$ in the ideal simulation, and the corresponding $p_{\mathrm{U}}$ is lower than 0.05, implying that the null hypothesis is rejected. Moreover, the STSR vanishes after mapping the empirical assemblage to the NSIT-compatible assemblage. These pieces of evidence suggest that, although the average teleportation fidelity exceeds the classical bound, it remains questionable to conclude that the experiment verifies the device's ability to generate genuine quantum information scrambling.

\section{Conclusion}\label{sec: conclusion}
In this work, we examine the reliability of the NSIT condition as a diagnostic tool for basis-dependent hardware errors for general quantum state transfer processes within the framework of spatio-temporal steering. Because realistic quantum experiments are performed with a finite number of measurement shots, statistical fluctuations alone can lead to apparent NSIT violations. To address this issue, we develop a statistical hypothesis-testing framework that provides the statistical evidence for intrinsic hardware imperfections. The corresponding null hypothesis states that the NSIT violation arises solely from statistical fluctuations. By combining the statistical behavior of $D$ under the null hypothesis with Chebyshev’s inequality, we derive a distribution-free upper bound on the $p$-value associated with an observed NSIT violation, enabling the null hypothesis to be rejected at a chosen significance level.

We demonstrate the utility of this approach through two examples. First, we perform a single-qubit quantum state transfer experiment on a superconducting processor and observe a consistent violation of the NSIT condition. By applying hypothesis test to the observed NSIT violation, we observe several instances in which the null hypothesis is rejected by a sufficiently small $p$-value. This provides the statistical evidence for a basis-dependent state-preparation error. Second, we consider a seven-qubit Hayden–Preskill teleportation protocol implemented on trapped-ion quantum computers provided by IonQ. Although the average teleportation fidelity on the IonQ Harmony device exceeds the classical threshold, we observed a statistically significant NSIT violation, indicating that basis-dependent hardware errors cannot be ruled out. Moreover, the STSR vanishes after mapping the empirical assemblage to the NSIT-compatible assemblage. These results suggest that fidelity-based indicators alone may be insufficient to certify genuine quantum information scrambling.

An important direction for future work is to extend the present analysis beyond the i.i.d. assumption that underlies most statistical treatments of finite-shot data~\cite{Zhang2013}. In realistic quantum devices, temporal drifts, correlated noise, and calibration fluctuations can introduce non-i.i.d. effects~\cite{Barrett2002, liang2019, Tabia_2025, Zhang2011} that may influence the interpretation of NSIT violations. Developing statistical tests that remain reliable in the presence of such correlations would further strengthen the robustness of spatio-temporal steering as a diagnostic tool for characterizing near-term quantum devices.

\section*{Acknowledgements}
The authors acknowledge fruitful discussions with Gelo Noel Tabia. The authors also acknowledge the Cloud Computing Center for Quantum Science \& Technology at NCKU (NSTC Grant No. 114-2119-M-006-003) for providing them a platform to implement the experiments. Y.-N.C. acknowledges the support of the National Center for Theoretical Sciences and the National Science and Technology Council, Taiwan (NSTC Grant No. 114-2112-M-006-015-MY3). T.-M.C. acknowledges the support of the Higher Education Sprout Project, Ministry of Education to the Headquarters of University Advancement at the National Cheng Kung University (NCKU).

\appendix 
\section{Configuration of the superconducting processor}\label{appx: qubit layout}
\begin{figure*}[!hbpt]
    \centering
    \includegraphics[width=0.8\textwidth]{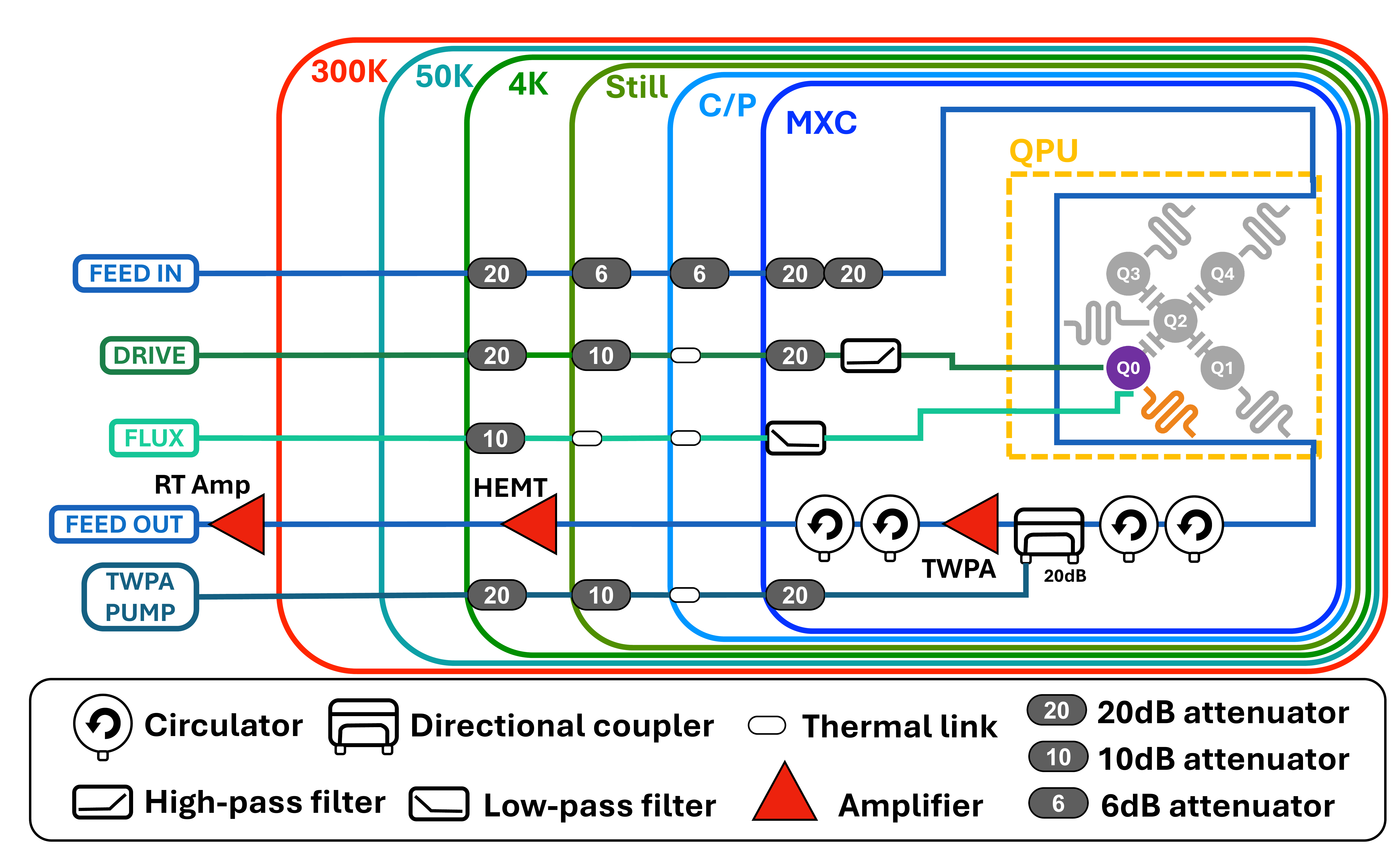}
    \caption{Cryogenic measurement setup and layout of the superconducting processor.}
    \label{fig: qubit layout}
\end{figure*}
\begin{figure*}[!hbpt]
    \includegraphics[width=\textwidth]{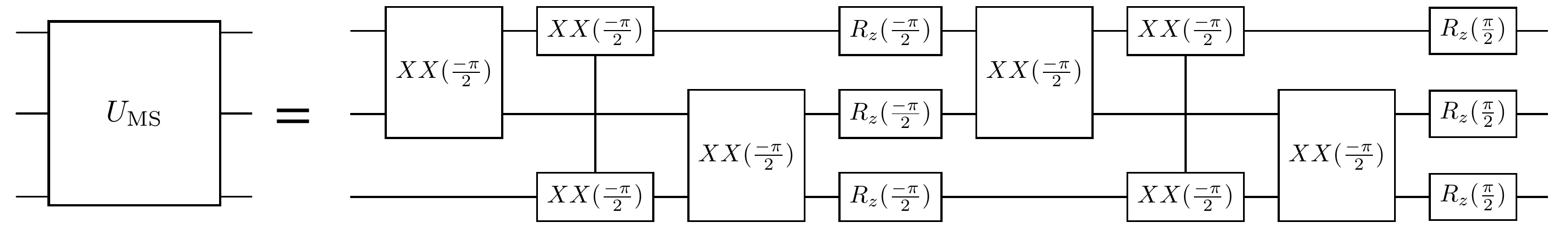}
    \caption{The circuit implementation of the three-qubit unitary gate $U_{\text{MS}}$.
    }
    \label{fig: scrambler}
\end{figure*}
The experiment in Sec.~\ref{sec: experiment 1} is performed in a Bluefors XLD1000 dilution refrigerator with a base temperature of 12 mK. The superconducting qubit device, a QuantWare Soprano 5-qubit processor, is housed within a multi-layered magnetic shield and thermally anchored to the mixing chamber stage. To preserve the cryogenic environment and mitigate thermal noise, the microwave input and flux control lines are appropriately attenuated and filtered across successive stages. On the output end, the weak dispersive readout signal is enhanced through a cascaded amplification chain to achieve a high signal-to-noise ratio. At the mixing chamber, the signal is first amplified by a Silent Waves Argo traveling-wave parametric amplifier (TWPA)~\cite{Planat2020}. It then passes through a cryogenic high-electron-mobility transistor (HEMT) amplifier at the 4 K stage, and finally a room-temperature amplifier. Standard cryogenic RF components, including a directional coupler and isolators, are distributed to ensure signal routing.

Room-temperature qubit control and readout are implemented using a Quantum Machines OPX1000 control platform. This controller is equipped with embedded microwave (MW-FEM) and low-frequency (LF-FEM) front-end modules. The MW-FEM supplies the microwave pulses for single-qubit gates and readout, while the LF-FEM provides the DC flux bias to set the qubit operating frequency.

A detailed schematic of the cryogenic wiring and component configurations is provided in Fig.~\ref{fig: qubit layout}, while the qubit parameters are summarized in Table ~\ref{table: qubit 0 stats}.

\begin{center}
\begin{table}
        \caption{
        Parameters of the qubit
}
    \label{table: qubit 0 stats}
    \renewcommand{\arraystretch}{1.5}
    \setlength{\tabcolsep}{14pt}
    \begin{tabular}{lc}
        \hline
        \hline
         & \textbf{$Q_0$}\\
        \hline
        {Readout frequency (GHz)}& {7.251}\\
        {Qubit frequency (GHZ)} & {4.513}\\
        {Anharmonicity (MHz)} & {-219}\\
        {$T_1 ( \mu \mathrm{s})$} & {15.3}\\
        {$T^*_2 (\mu \mathrm{s})$} & {9.8}\\
        {Readout fidelity} & {0.843}\\
        \hline
        \hline
    \end{tabular}
\end{table}
\end{center}

\section{The circuit decomposition of the maximally scrambling unitary}\label{appx: scrambler circuit}
We implement the maximally scrambling unitary $U_{\mathrm{MS}}$ by the combination of the local single-qubit rotation gate $R_z(\theta)$ and the two-qubit entangling gate $XX(\theta)$, as shown in Fig.~\ref{fig: scrambler}. These gates are defined as
\begin{equation}
    R_z(\theta) = e^{ i\frac{\theta}{2}\sigma_z}
\end{equation}
and 
\begin{equation}
    XX(-\pi/2) = e^{ i\frac{\pi}{4}(\sigma_x\otimes \sigma_x)}. 
\end{equation}
The degree of scrambling is controlled by tuning the parameter $\theta$ of the $R_z(\theta)$ gates. To realize the maximally scrambling unitary, $\theta$ is set to $\pi/2$ in the first block and $-\pi/2$ in the second block, as indicated in Fig.~\ref{fig: scrambler}.

\bibliography{ref.bib}
\end{document}